\documentclass[12pt]{article}

\newcommand{\sect}[1]{\setcounter{equation}{0}\section{#1}}

\newcommand{\eq}{\begin{equation}}
\newcommand{\eqa}{\begin{eqnarray}}
\newcommand{\en}{\end{equation}}
\newcommand{\ena}{\end{eqnarray}}
\newcommand{\enn}{\nonumber \end{equation}}

\def\sk{\vskip .4cm}
\def\noi{\noindent}
\def\om{\omega}
\def\al{\alpha}
\def\be{\beta}
\def\ga{\gamma}
\def\Ga{\Gamma}

\let \part\partial

\def\unmezzo{{1 \over 2}}
\def\epsi{\varepsilon}
\def\we{\wedge}

\def\de{\delta}

\def\tv{{\bf t}}

\def\part{\partial}

\def\sk{\vskip .4cm}

\def\noi{\noindent}

\def\X0{X^0}

\def\om{\omega}

\def\al{\alpha}
\def\ga{\gamma}

\def\unmezzo{{1 \over 2}}
\def\epsi{\varepsilon}

\def\psib{{\bar \psi}}

\def\rhob{\bar\rho}

\def\we{\wedge}

\def\de{\delta}

\def\tbo{{\bf t}}

\def\Dcal{{\cal D}}
\def\Rcal{{\cal R}}

\def\square{{\,\lower0.9pt\vbox{\hrule \hbox{\vrule height 0.2 cm
\hskip 0.2 cm \vrule height 0.2 cm}\hrule}\,}}

\def\epsilonbar{{\bar \epsilon}}

\begin{document}

\begin{titlepage}
\rightline{DISTA-UPO/06}
\rightline{April 2006}
 \vskip 2em
\begin{center}{\bf   EXTENDED LIE DERIVATIVES AND A NEW FORMULATION OF D=11 SUPERGRAVITY}
\\[3em]
Leonardo Castellani\\[1em] {\sl Dipartimento di Scienze e
Tecnologie Avanzate\\ Universit\`a del Piemonte Orientale\\ Via
Bellini 25/G, 15100 Alessandria, Italy}\\[.5 em] {\sl and\\[.5em]
Istituto Nazionale di Fisica Nucleare\\ Via Giuria 1, 10125
Torino, Italy.}
  \\[3em]

\end{center}

\begin{abstract}
Introducing an extended Lie derivative along the dual of $A$, the
three form field of d=11 supergravity, the full diffeomorphism
algebra of  d=11 supergravity is presented.  This algebra suggests
a new formulation of the theory, where the three-form field $A$ is
replaced by bivector $B^{ab}$, bispinor $B^{\alpha\beta}$, and
spinor-vector $\eta^{a\beta}$ one-forms. Only the bivector
one-form $B^{ab}$ is propagating, and carries the same degrees of
freedom of the three-form in the usual formulation, its curl
$\Dcal_{[\mu} B^{ab}_{\nu]}$ being related to the
 $F_{\mu \nu a b}$ curl of the three-form.
The other one-forms are auxiliary, and the transformation rules
on all the fields close on the equations of motion of d=11
 supergravity.

\end{abstract}

\vskip 6cm \noi \hrule \vskip.2cm \noi {\small
leonardo.castellani@mfn.unipmn.it}

\end{titlepage}

\newpage
\setcounter{page}{1}

\sect{Introduction}

Supergravity in eleven dimensions \cite{d11} (a particular limit
of M-theory, for a review see for ex. \cite{mtheory}), can be
formulated within the framework of free differential algebras
(FDA's)  \cite{fda0,fda1,fda2,gm2,gm3,fda3}, a generalization of
Lie algebras which include $p$-form potentials.  In this framework
the 3-form of $d=11$ supergravity acquires an algebraic
interpretation.

In preceding papers \cite{fda3,d11al}  it was shown how FDA's can
be related to ordinary Lie algebras via extended Lie derivatives:
the Lie algebra underlying the FDA of $d=11$ supergravity was
identified in \cite{d11al}, and found to coincide with an algebra
discussed in 1989 \cite{BergSez} in the context of supermembranes.

In Section 2 we recall briefly the FDA formulation of $d=11$
supergravity of \cite{fda1}. In Section 3 we present the full
diffeomorphism algebra of $d=11$ supergravity on the FDA
``manifold", which encodes all the symmetries of the theory. In
Section 4 a new formulation is proposed, where the 3-form $A$ is
replaced by 1-form potentials with (couples of) Lorentz vector and
spinor indices.

For a resum\'e on FDA's and the notion of extended Lie derivative
along duals of $p$-forms we refer to  \cite{d11al}. Here  the
general theory is applied to the FDA of $d=11$ supergravity.

\sect{The FDA of d=11 supergravity, Bianchi identities, field
equations and transformation rules}

The FDA structure \cite{fda1} is contained in the following
curvature definitions:
 \eqa
 & & R^{ab}=d\om^{ab}-\om^{ac} \om^{cb} \nonumber\\
 & & R^a=dV^a-\om^{ab}V^b-{i\over 2}\psib \Gamma^a \psi
  \equiv \Dcal V^a -{i\over 2}\psib \Gamma^a \psi\nonumber\\
 & & \rho=d\psi - {1\over 4} \om^{ab} \Gamma^{ab} \psi \equiv \Dcal \psi \nonumber\\
 & & R(A)=dA - {1\over 2} \psib \Gamma^{ab} \psi V^a V^b \label{FDAd11}
 \ena

\noi The Bianchi identities are obtained by taking the exterior
derivative of (\ref{FDAd11}):
 \eqa
 & &  dR^{ab} + R^{ac} \om^{cb}-\om^{ac} R^{cb} \equiv  {\cal D} R^{ab}=0\nonumber\\
 & & {\cal D}R^a+R^{ab}V^b- i \psib \Gamma^a \rho=0\nonumber\\
 & & {\cal D} \rho+{1\over 4} R^{ab} \Gamma^{ab} \psi  =0\nonumber\\
 & & dR(A)-\psib \Gamma^{ab} \rho V^a V^b +\psib \Gamma^{ab} \psi R^a V^b=0 \label{BId11}
 \ena

The superPoincar\'e curvatures $R^{ab},R^a,\rho$ (respectively the
Lorentz curvature, the torsion and the gravitino curvature) are
2-forms, and $R(A)$ is a 4-form. These can be expanded on a
superspace basis spanned by the vielbein $V^a$ and the gravitino
$\psi$. The group-geometric method of \cite{gm1,gm2,gm3} requires
the Ansatz that all  ``exterior" components of the curvatures be
expressed in terms of the ``spacetime" ones, spacetime meaning
along the $V^a$ vielbeins only.

The Bianchi {\sl identities} then become {\sl equations} for the curvatures, whose solution
\cite{fda1} is:
 \eqa
 R^{ab}&=& R^{ab}_{~~cd} V^c V^d
 +i~(2\rhob_{c[a}\Gamma_{b]}-\rhob_{ab}\Gamma_c)~\psi V^c \nonumber \\
 & & ~~~~~~~~+
 F^{abcd} ~\psib \Gamma^{cd} \psi+ {1 \over 24} F^{c_1c_2c_3c_4}~
 \psib \Gamma^{abc_1c_2c_3c_4} \psi \label{Rab}\\
  R^a &=& 0 \label{Ra}\\
 \rho & =& \rho_{ab} V^a V^b + {i \over 3} (F^{ab_1b_2b_3}
 \Gamma^{b_1b_2b_3}-{1\over 8}F^{b_1b_2b_3b_4}
 \Gamma^{ab_1b_2b_3b_4})~ \psi V^a \label{rho}\\
 R(A)&=& F^{a_1...a_4}V^{a_1}V^{a_2}V^{a_3}V^{a_4}\label{RA}
 \ena

 \noi where the spacetime components $R^{ab}_{~~cd},\rho_{ab},  F^{a_1...a_4}$
 satisfy the well known
  propagation equations (Einstein, gravitino and Maxwell equations):
 \eqa
 & & R^{ac}_{~~bc} - \unmezzo \de^a_b R = 3~ F^{ac_1c_2c_3}
 F^{bc_1c_2c_3}-{3 \over 8}~\de^a_b~F^{c_1...c_4}
 F^{c_1...c_4} \label{Einstein}\\
 & & \Gamma^{abc} \rho_{bc}=0 \label{gravitino}\\
 & & \Dcal_a F^{ab_1b_2b_3} - {1 \over 2 \cdot 4!\cdot 7!}~
 \epsilon^{b_1b_2b_3a_1...a_8} ~F^{a_1...a_4} F^{a_5...a_8} =0 \label{Maxwell}
 \ena

In the group geometric formulation the symmetries gauged by the
superPoincar\'e fields $V^a$, $\om^{ab}$ and $\psi$ are seen as
diffeomorphisms on the ``FDA manifold", generated by the Lie
derivative along the tangent vectors $t_a,t_{ab},\tau$ dual to
these one-form fields. Thus, setting $\epsi= \epsi^a t_a +
\epsi^{ab} t_{ab}+\epsilon \tau$, the transformation rules under
local supertranslations and Lorentz rotations are generated by the
Lie derivative
 \eq
 \ell_{\epsi} \equiv d~ i_{\epsi} + i_{\epsi} ~d
\label{Liederivative}
\en

\noi Explicitly
\eqa & &  \de V^a = \ell_{\epsi} V^a= \Dcal
\epsi^a + \epsi^{ab} V^b + i \epsilonbar \Ga^a \psi
\label{LieonV}\\ & &  \de \om^{ab}= \ell_{\epsi}\om^{ab}=\Dcal
\epsi^{ab} + 2 R^{ab}_{~~cd} ~\epsi^c V^d
 +i~(2\rhob_{c[a}\Gamma_{b]}-\rho_{ab}\Gamma_c)~(\epsilon V^c - \psi \epsi^c)- \nonumber \\
 & & ~~~~~~~~~~~~~~~~~
-2 F^{abcd} ~\psib \Gamma^{cd} \epsilon- {1 \over 12} F^{c_1c_2c_3c_4}~
 \psib \Gamma^{abc_1c_2c_3c_4} \epsilon  \label{Lieonom}\\
& & \de \psi =  \ell_{\epsi}\psi= \Dcal \epsilon + {i\over 4}
\epsi^{ab} \Ga_{ab} \psi +
 2 \rho_{ab} \epsi^a V^b + \nonumber\\
 & & ~~~~~~~~~~~~~~~~~+ {i \over 3} (F^{ab_1b_2b_3}
 \Gamma^{b_1b_2b_3}-{1\over 8}F^{b_1b_2b_3b_4}
 \Gamma^{ab_1b_2b_3b_4})~ (\epsilon V^a - \psi \epsi^a) \label{Lieonpsi}\\
 & & \de A=\ell_{\epsi} A=-\psib \Gamma^{ab} \epsilon ~ V^a V^b +
 \psib \Gamma^{ab} \psi~ \epsi^a V^b+
 4 F^{a_1...a_4}\epsi^{a_1}V^{a_2}V^{a_3}V^{a_4} \label{LieonA}
\ena

\noi where the exterior derivatives on the fields have been
expressed in terms of the curvatures (\ref{FDAd11}), and the
solutions (\ref{Rab})-(\ref{RA}) have been used. The closure of
these transformations is then equivalent to the propagation
equations (\ref{Einstein})-(\ref{Maxwell}), as is usual in locally
supersymmetric theories.

\sect{The algebra of diffeomorphisms on the d=11 supergravity FDA ``manifold"}

On a soft group manifold, i.e. a manifold whose vielbeins $\mu^A$
have in general nonvanishing curvatures
 \eq
 R^A = d\mu^a + \unmezzo
C^C_{AB} \mu^A \mu^B, \label{RLie}
 \en
 \noi the algebra of diffeomorphisms
 is given by the commutators of Lie derivatives:
 \eq
 \left[ \ell_{ \epsi^A_1 \tbo_A},\ell_{
\epsi^B_2 \tbo_B} \right] = \ell_{ \left[ \epsi^A_1 \partial_A
\epsi^C_2 - \epsi^A_2 \partial_A \epsi^C_1 - 2 \epsi^A_1 \epsi^B_2
\Rcal^C_{AB} \right] \tbo_C} \label{aldiffLie}
\en

\noi where $\tbo_A$ are the tangent vectors dual to the one-forms
$\mu^A$, and
 \eq \Rcal^C_{AB} \equiv R^C_{AB}-\unmezzo C^C_{AB}
\label{Rcal}
\en
involves the curvature components on the vielbein basis and the group structure constants.
The closure of the algebra requires the Bianchi identities
\eq
\part_{[B} \Rcal^A_{CD]}+2~\Rcal^A_{E[B} \Rcal^E_{CD]} =0
\label{BianchiRcal}
\en

On a soft ``FDA manifold" , the algebra of diffeomorphisms
includes also the diffeomorphisms in the $p$-form directions,
generated by an extended Lie derivative $\ell_{\epsi \tbo} $,
where $\tbo$ is a ``tangent vector" dual to the $p$-form, and
$\epsi$ is a $p-1$ form parameter  \cite{fda3,d11al}.

In the case of $d=11$ supergravity all the local symmetries of the
theory are given by the following FDA diffeomorphism algebra:
 \eqa
 & &  \left[ \ell_{ \epsi^A_1 \tbo_A},\ell_{
\epsi^B_2 \tbo_B} \right] = \ell_{ \left( \epsi^A_1 \partial_A
\epsi^C_2 - \epsi^A_2 \partial_A \epsi^C_1 - 2 \epsi^A_1 \epsi^B_2
\Rcal^C_{AB} \right) \tbo_C} + \nonumber\\ & &  +  \ell_{\left(
-\epsilonbar_1 \Ga^{ab} \epsilon_2 V^a V^b -  \epsi_1^a \epsi_2^b~
\psib \Ga^{ab} \psi
  + 2 \epsi^a_2 \epsilonbar_1~ \Ga^{ab} \psi  V^b-
  2 \epsi^a_1 \epsilonbar_2 ~\Ga^{ab} \psi  V^b -12 \epsi^{a}_1
 \epsi^{b}_2~ F^{abcd} V^{c}V^{d} \right) \tbo} \label{FDAdiff1}\\
 & & \left[ \ell_{ \epsi^A  \tbo_A},\ell_{
\epsi \tbo} \right] =  \ell_{ \zeta \tbo} \label{FDAdiff2}\\ & &
\left[ \ell_{ \epsi_1 \tbo},\ell_{ \epsi_2 \tbo} \right] =0
\label{FDAdiff3}\ena

\noi where the indices {\small {\sl A,B,C...}}  run on the Lie
algebra directions (corresponding to the vielbein, gravitino and
spin connection one-forms), i.e. {\small {\sl A}}
 $= a, \al, ab$(Lorentz). The quantities $ \Rcal^C_{AB}$ (cf.
 (\ref{RLie})) are given by the solutions for $R^{ab}, R^a, \rho$
 in (\ref{Rab})-(\ref{rho}) and by the superPoincar\'e structure
 constants encoded in the first three lines of (\ref{FDAd11}).
 The two-form parameters $\epsi$ and $\zeta$ in (\ref{FDAdiff2})
 are
\eqa
 & & \epsi \equiv \epsi_{ab}V^aV^b+2\epsi_{a\be}V^a
 \psi^\be+\epsi_{\al\be}\psi^\al \psi^\be \label{epsi}\\
& & \zeta \equiv  \epsi^A (\Dcal_A \epsi_{cd}) V^c V^d-2 \epsi^c
(\Dcal \epsi_{cd}) V^d + 2i\epsi_{cd} \epsilonbar \Ga^c \psi V^d
+i \epsi_{cd} \epsi^d \psib \Ga^c \psi + \nonumber\\
 & &~~~~~+  2 \epsi^A (\Dcal_A \epsi_{c\al}) V^c \psi^\al-2 \epsi^c
 (\Dcal \epsi_{c\al})\psi^\al-2 \epsi^\al (\Dcal \epsi_{c\al})V^c+
 2i  \epsi_{c\al} \epsilonbar \Ga^c \psi \psi^\al + \nonumber\\
 & &~~~~~+ i \epsi_{c\al} \epsi^\al \psib \Ga^c \psi - 2 \epsi_{c\al}
  \epsi^c \rho^\al - 4\epsi_{c\al}(\rho^\al_{Ab}\epsi^A V^b +\rho^\al_{A \beta}\epsi^A
  \psi^\beta )V^c+ \nonumber\\
  & & ~~~~~+ \epsi^A (\Dcal_A \epsi_{\al\be}) \psi^\al \psi^\be-2\epsi^\al (\Dcal
  \epsi_{\al\be})  \psi^\be+2\epsi_{\al\ga}(\rho^\al_{Ab}\epsi^A V^b +\rho^\al_{A \beta}\epsi^A
  \psi^\beta )\psi^\ga - \nonumber\\
  & & ~~~~~-2 \epsi_{\al\be} \epsi^\al\rho^\be \label{zeta}
 \ena

 \noi ($\epsi_{ab}$ here not to be confused with the Lorentz
 rotation parameter of Section 2).
 The Lie derivative along $\tbo$, the dual of the
 three form $A$, is a particular case of the extended Lie derivatives along $p$-forms
 $B^i$  ($i$ being a $G$-representation index) introduced
 in \cite{fda3,d11al}, the fields in this general setting being the $G$ Lie algebra
 one-forms $\mu^A$ supplemented by the $p$-form $B^i$. The extended Lie derivative is
 given by
 \eq
 \ell_{\epsi^i \tv_i} \equiv i_{\epsi^i \tv_i}d + d~
i_{\epsi^i \tv_i} \label{newLie}
 \en

\noi the contraction operator  $i_{\epsi^i \tv_i}$ being defined
by its action on a generic form $\om = \om_{i_1...i_n A_1...A_m}
B^{i_1} \we ... B^{i_n} \we \mu^{A_1} \we ... \mu^{A_m}$ as
 \eq
i_{\epsi^j \tv_j} \om = n~ \epsi^j \om_{j i_2...i_n A_1...A_m}
B^{i_2} \we ... B^{i_n} \we \mu^{A_1} \we ... \mu^{A_m}
\label{newcontraction}
\en
\noi  where $\epsi^j$  is a $(p-1)$-form. Thus the contraction
operator still maps $p$-forms into $(p-1)$-forms. Note that i)
$i_{\epsi^j \tv_j}$ vanishes on forms that do not contain at least
one factor $B^i$; ii) the extended Lie derivative commutes with
$d$ and satisfies the Leibnitz rule.

Returning to the FDA of $d=11$ supergravity, since $A$ is a
three-form in the identity representation of the superPoincar\'e
Lie algebra, parameters in the extended Lie derivative (along the
dual $\tbo$ of $A$) are 2-forms carrying no representation index,
and are explicitly given for the algebra of FDA diffeomorphisms in
 (\ref{zeta}) and (\ref{epsi}).

The action of the extended Lie derivative on the basic fields is
simply:
\eq
 \ell_{\epsi \tbo} \mu^A =0,~~~~\ell_{\epsi \tbo} A = d\epsi
 \en

\noi with $\mu^A = V^a, \om^{ab}, \psi$.
 Using these rules together with the variations
  (\ref{LieonV})-(\ref{LieonA}) (generated by
 the usual Lie derivative) leads to the diffeomorphism algebra
 given in eq.s (\ref{FDAdiff1})-(\ref{FDAdiff3}). As discussed
 in ref. \cite{d11al} for the general case, the algebra of FDA
 diffeomorphisms closes provided the FDA Bianchi identities hold.
 Therefore, if we use in (\ref{FDAdiff1}) and
 (\ref{FDAdiff2}) the solutions (\ref{Rab})-(\ref{rho})
 for the curvatures, the algebra (\ref{FDAdiff1})-(\ref{FDAdiff3})
 closes on the $d=11$ field equations
  (\ref{Einstein}) - (\ref{Maxwell}).

  Note that the commutator of two ordinary Lie derivatives,
  computed on the 3-form $A$, {\sl does not close} on the
  usual (\ref{aldiffLie}) diffeomorphism algebra, but
  develops an extra piece, i.e. the second line in the ``extended"
  diffeomorphism algebra of (\ref{FDAdiff1}), containing the
  extended Lie derivative.

 \sect{A new formulation of D=11 supergravity}

 The idea is to reinterpret the extended Lie derivative
  $\ell_{\epsi \tbo} $ of the
 $d=11$ FDA in terms of ordinary Lie derivatives along
 new tangent vectors $\tbo_{ab}$,  $\tbo_{a\be}$, $\tbo_{\al\be}$,
 via the following identification:
 \eq
   \ell_{\epsi \tbo}=\ell_{\epsi^{ab} V^a V^b \tbo+
   \epsi^{a\be} V^a \psi^\be \tbo+ \epsi^{\al\be} \psi^\al \psi^\be \tbo}
    \equiv  \ell_{\epsi^{ab}  \tbo_{ab}+
   \epsi^{a\be}\tbo_{a\be}+ \epsi^{\al\be} \tbo_{\al\be}}
   \en
\noi The 0-forms $ \epsi^{ab}$,  $\epsi^{a\be}$, $\epsi^{\al\be}$,
i.e. the coefficients of the expansion on the superspace basis of
the 2-form parameter
 $\epsi$ in the extended Lie derivative, are reinterpreted as parameters of
  ordinary Lie derivatives
along the new tangent vectors $\tbo_{ab}=V^a V^b \tbo$,
 $\tbo_{a\be}=V^a \psi^\be \tbo$, $\tbo_{\al\be}= \psi^\al \psi^\be \tbo$.

This is possible when the set
 \eq
 \ell_{\epsi^A \tbo_A}, \ell_{\epsi^{ab} \tbo_{ab}},
 \ell_{\epsi^{a\be} \tbo_{a\be}}, \ell_{\epsi^{a\be} \tbo_{a\be}} \label{Lieall}
 \en

\noi closes on a diffeomorphism algebra similar to the one in
(\ref{aldiffLie}), i.e. a diffeomorphism algebra of an ordinary
group manifold.  If this is the case the new operators can be seen as  {\sl
bona fide} Lie derivatives, generating ordinary diffeomorphisms
along new directions.

Now (\ref{FDAdiff1}) indeed is of the form (\ref{aldiffLie}), and
the extra piece on the right hand side simply defines new
curvatures and structure constants in $\Rcal^C_{AB}$ of
(\ref{Rcal}). However the other commutations (\ref{FDAdiff2})
contain terms with exterior (covariant) derivatives of the
parameters $ \epsi^{ab}$, $\epsi^{a\be}$, $\epsi^{\al\be}$,  not
amenable to the form of the derivative terms in (\ref{aldiffLie}).
These parameters (associated with the new directions) will
therefore be taken to be covariantly constant in the arguments
that follow. This is the price to pay if we want to interpret the
$d=11$ diffeomorphism algebra (\ref{FDAdiff1})-(\ref{FDAdiff3}) as
an algebra of ordinary Lie derivatives.

In other words: the algebra (\ref{FDAdiff1})-(\ref{FDAdiff3}) with
$\Dcal \epsi_{cd}=\Dcal \epsi_{c\al}=\Dcal \epsi_{\al\be}=0$ can
be considered the diffeomorphism algebra of a manifold, whose
vielbeins are
 $V^a$, $\om^{ab}$, $\psi$,  $B^{ab}$,  $B^{\alpha\beta}$, and
$\eta^{a\beta}$  (the last three being the vielbeins dual to the
tangent vectors $\tbo_{ab}$,
 $\tbo_{a\be}$, $\tbo_{\al\be}$).

 Comparing (\ref{FDAdiff1})-(\ref{FDAdiff3}) with the general form
(\ref{aldiffLie}) we deduce the new curvature components that
satisfy the Bianchi identities (\ref{BianchiRcal}) implied by
(\ref{aldiffLie}):  $ R^{ab}, \rho, R^a$ remain unchanged as given
in eq.s (\ref{Rab})-(\ref{rho}), whereas the solutions for the
curvatures $T^{ab}, T^{\al\be}, \Sigma^{a\be}$
 of the new potentials $B^{ab}$,  $B^{\alpha\beta}$,
$\eta^{a\beta}$ are
 \eqa
 & & T^{ab}=24~ F^{ab}_{~~cd} V^c V^d -{3 \over 4} \rho^{\de}_{[ab}
  \eta_{c]\de} V^c - {1 \over 4} \rho^{\de}_{\al[a}
  \eta_{b]\de} \psi^{\al} - {1 \over 2} \rho_{ab}^\ga \psi^\de
  B_{\ga\de} \label{solTab}\\
  & & T^{\al\be} = {1 \over 4} ~\rho^\ga_{a\{ \al} B_{\be \} \ga}
  V^a  \label{solTalbe}\\
  & & \Sigma^{b\be} = -i \rho^{\{\al}_{ab} B^{\be\}\al} V^a - {i
  \over 2} \rho^\de_{\be [ a}\eta^\de_{b]} V^a - {i
  \over 2} \rho^\ga_{b \{ \al} B_{\be \} \ga} \psi^\al
  \label{solsigma}
  \ena

 \noi (all contractions Lorentz invariant, position of indices not relevant).

Using the standard formula for the variation of group manifold vielbeins $\mu^A$
 under diffeomorphisms:
 \eq
 \de \mu^A = d \epsi^A - 2 \Rcal^A_{~BC} \mu^B \epsi^C
 \label{varmu}
 \en

\noi we find the variations of the potentials $B^{ab}$,
$B^{\alpha\beta}$, and $\eta^{a\beta}$:
 \eqa
  & & \de B^{ab} =  \psib \Ga^{ab} \epsilon - 48 F^{abcd} V^c \epsi^d
 +{3\over 2} \rho^\de_{[ab} (\epsi_{c]}^{~\de} V^c -  \epsi^c \eta_{c]}^{~\de})
+\nonumber\\ & & ~~~~~~ + {1\over 2} \rho^\de_{\al [a} (
\epsi_{b]}^{~\de} \psi^\al
 +  \epsi^\al \eta^{~\de}_{b]}) + \rho^\ga_{ab}(\epsi^\de
 B^{\ga\de} - \epsi^{\ga\de} \psi^\de)  \label{LieonBab}\\
 & & \de B^{\al\be} =  2
 (C\Ga_{ab})^{\al\be} V^a \epsi^b - {3i \over 2} (C\Ga_c)_{ \{ \al \be}
 (\epsi^\ga \eta_{\ga \} }^{~c} + \epsi_{\ga \} }^{~c}
 \psi^\ga) +  \nonumber\\
  & & ~~~~~~ + i(C\Ga_c)^{\al\be} (\epsi^d B^{cd}
 - \epsi^{cd} V^d) - {1\over 2} \rho^\ga_{a\{\al}
 (\epsi^\ga_{~\be\}} V^a - \epsi^a B^\ga_{~\be\}})\\
 & & \de \eta^{b\be} = 2 (C\Ga_{ab})_{\de\be}
 (V^a \epsi^\de - \psi^\de \epsi^a) + 2i (C\Ga^c)_{\al\be}
 (\psi^\al \epsi^{bc} - B^{bc} \epsi^\al) +\nonumber\\
 & & ~~~~~~ + 2i \rho^{\{\ga}_{ab} (\epsi^{\be\}\ga} V^a - \epsi^a
 B^{\be\}\ga}) + i \rho^\de_{\be [a} (\epsi_{b]}^{~\de} V^a -
 \epsi^a \eta_{b]}^{~\de}) \nonumber \\
  & & ~~~~~~ + i \rho^\ga_{b \{ \al}
 (\epsi_{\be \} \ga} \psi^\al - \epsi^\al B_{\be \} \ga})
 \label{Lieoneta}
 \ena

 \noi having used $\Dcal
\epsi^{ab}=  \Dcal \epsi^{\al\be} = \Dcal \epsi^{b\be} =0$
 in (\ref{varmu}). The variations of  $V^a$,
$\om^{ab}$, $\psi$ are unchanged and given in
(\ref{LieonV})-(\ref{Lieonpsi}).

 In summary: the new formulation of $d=11$ supergravity proposed
 here contains the fields:
 \eq
V^a, ~\om^{ab}, ~\psi,
 ~B^{ab}, ~B^{\alpha\beta}, ~\eta^{a\beta} \label{newfields}
 \en

 \noi The transformation rules of these fields, given in
 (\ref{LieonV})-(\ref{Lieonpsi}) and (\ref{LieonBab})-(\ref{Lieoneta})
 close under the same conditions necessary for the closure of
 the algebra (\ref{FDAdiff1})-(\ref{FDAdiff3}), since it is just
 a reformulation of this algebra. The only "spurious" element in this
 reformulation is the fact that some of the parameters
 (i.e.  $ \epsi^{ab}$, $\epsi^{a\be}$, $\epsi^{\al\be}$ )
 must be taken to be covariantly constant. Even in this case
 the algebra (\ref{FDAdiff1})-(\ref{FDAdiff3}) closes only
 provided the Bianchi identities (\ref{BId11}) hold, which implies
 the $d=11$ field equations (\ref{Einstein})-(\ref{Maxwell}).
 Thus the closure of the transformation rules on the fields
 (\ref{newfields}) requires the $d=11$ field equations, a situation
 analogous to the one in type IIB supergravity \cite{2b}.

Finally, we can relate the covariant curl $\Dcal_{[\mu}
B^{ab}_{\nu]}$
 to the curl of the three-form $F_{\mu \nu a b}$: indeed
 the curvature of $B^{ab}$, according to the definition
 (\ref{RLie}), reads
 \eq
 T^{ab} =  \Dcal B^{ab} - {1 \over 2} \psib \Ga^{ab} \psi
 \label{defTab}
 \en
\noi (see also \cite{d11al})  where we have used the structure
constants deduced by recasting the diffeomorphism algebra
(\ref{FDAdiff1})-(\ref{FDAdiff3}) in the form (\ref{aldiffLie}).
Comparing the $V^c V^d$ components of the definition of $T^{ab}$
(\ref{defTab}) and its solution (\ref{solTab}) yields
 \eq
  (\Dcal B^{ab})_{cd} = 24 F^{ab}_{~~cd}
   \en
\noi $(\Dcal B^{ab})_{cd}$ being the $V^c V^d$ components of the
two-form $\Dcal B^{ab}$. The other fields $B^{\al\be}$,
$\eta^{a\be}$ are auxiliary: their curvature solutions, given
respectively in (\ref{solTalbe}) and (\ref{solsigma}), have no
spacetime ($VV$)-components, and the external components only
contain the gravitino curvature.

In conclusion, we have found a set of transformation rules on the
 dynamical fields $V^a, ~\om^{ab}, ~\psi,
 ~B^{ab}$ and auxiliary fields $B^{\alpha\beta}, ~\eta^{a\beta}$
 that close on the (usual) field equations of $d=11$
 supergravity, $F^{abcd}$ being now related to the curl of
 $B^{ab}$.

\sk
 {\bf Acknowledgements}
 \sk

This work is partially supported by the European Community's Human
Potential Program under contract MRTN-CT-2004-005104 and by the
Italian MIUR under contract PRIN-2003023852 on the project
``Superstrings, branes and fundamental interactions".

\vfill\eject

\begin{thebibliography}{99}

\bibitem{d11} E. Cremmer and B. Julia, {\sl Supergravity theory in eleven dimensions}, Phys. Lett. {\bf B76} (1978)409;
{\sl The SO(8) supergravity}, Nucl. Phys. {\bf B159} (1979)141.

\bibitem{mtheory} P.K. Townsend, {\sl Four lectures on M theory},
Proceedings of ``High energy physics and cosmology", 385-438,
Trieste 1996, hep-th/9612121;  {\sl M theory from its
superalgebra}, Proceedings of ``Strings, branes and dualities",
141-177, Cargese 1997, hep-th/9712004.

\bibitem{fda0} D. Sullivan, {\sl Infinitesimal computations in
topology}, Bull. de L' Institut des Hautes Etudes Scientifiques,
Publ. Math. {\bf 47} (1977).

\bibitem{fda1} R. D' Auria and P. Fr\'e, {\sl Geometric supergravity
in D=11 and its hidden supergroup}, Nucl. Phys. {\bf B201} (1982)
101;

\bibitem{fda2}
 L. Castellani, P. Fr\'e, F. Giani, K. Pilch and P. van
Nieuwenhuizen,  {\sl Gauging of D=11 supergravity ?} Ann. Phys.
{\bf146} (1983) 35; P. van Nieuwenhuizen, {\sl Free graded
differential algebras}
 in: Group theoretical methods in physics, Lect. Notes in Phys. 180
(Springer, Berlin, 1983).

\bibitem{gm2} L. Castellani, R. D'Auria and P. Fr\'e, {\sl Supergravity and
superstrings: a geometric perspective}, World Scientific,
Singapore 1991.

\bibitem{gm1} Y. Ne'eman and T. Regge,  {\sl Phys. Lett.} {\bf B74} (1978) 31;
{\sl Riv. Nuovo Cimento} {\bf 1} (1978) 5; A. D' Adda, R. D'
Auria, P. Fr\'e and T. Regge, {\sl Riv. Nuovo Cimento} {\bf 3}
(1980) 6; R. D' Auria, P. Fr\'e and T. Regge, {\sl Riv. Nuovo
Cimento} {\bf 3} (1980) 12.

\bibitem{gm3} L. Castellani, {\sl Group geometric methods in supergravity and superstring theories},
 Int. J. Mod. Phys. {\bf A7} (1992) 1583.

\bibitem{fda3} L. Castellani and A. Perotto, {\sl Free differential
algebras, their use in field theory and dual formulation}, Lett.
Math. Phys. {\bf 38} (1996) 321, hep-th/9509031.

\bibitem{d11al}
  L.~Castellani,
  {\sl Lie derivatives along antisymmetric tensors, and the M-theory
  superalgebra }, hep-th/0508213.

  \bibitem{BergSez} E. Bergshoeff, E. Sezgin,
  {\sl New spacetime superalgebras and their Kac-Moody extension},
   Phys. Lett. {\bf B232} (1989) 96.


\bibitem{2b} M.B. Green and J.H Schwarz, Phys. Lett. {\bf 109B} (1982)
 444; {\bf 122B} (1983) 43.

\end{thebibliography}
\end{document}